# RDPP-TD: Reputation and Data Privacy-Preserving based Truth Discovery Scheme in Mobile Crowdsensing

Lijian Wu[a], Weikun Xie[a], Wei Tan[a], Tian Wang[b], Houbing Herbert Song[c], *Fellow, IEEE*, Anfeng Liu[a,*]

***Abstract*—Truth discovery (TD) plays an important role in Mobile Crowdsensing (MCS). However, existing TD methods, including privacy-preserving TD approaches, estimate the truth by weighting only the data submitted in the current round, which often results in low data quality. Moreover, there is a lack of effective TD methods that preserve both reputation and data privacy. To address these issues, a Reputation and Data Privacy-Preserving based Truth Discovery (RDPP-TD) scheme is proposed to obtain high-quality data for MCS. The RDPP-TD scheme consists of two key approaches: a Reputation-based Truth Discovery (RTD) approach, which integrates the weight of current-round data with workers' reputation values to estimate the truth, thereby achieving more accurate results, and a Reputation and Data Privacy-Preserving (RDPP) approach, which ensures privacy preservation for sensing data and reputation values. First, the RDPP approach, when seamlessly integrated with RTD, can effectively evaluate the reliability of workers and their sensing data in a privacy-preserving manner. Second, the RDPP scheme supports reputation-based worker recruitment and rewards, ensuring high-quality data collection while incentivizing workers to provide accurate information. Comprehensive theoretical analysis and extensive experiments based on real-world datasets demonstrate that the proposed RDPP-TD scheme provides strong privacy protection and improves data quality by up to 33.3%.**

## I. INTRODUCTION

Mobile Crowdsensing is one of the large-scale data collection paradigms that leverages ubiquitous workers to sense and collect massive amounts of data [1], [2], [3], enabling the construction of numerous data-based applications or services [4], [5]. Benefiting from the acquisition of large-scale data, artificial intelligence services based on "large models + big data" training have shown explosive development [6], [7]. MCS finds applications in diverse domains, including air quality traffic detection [8], [9], and street view collection [10]. Generally, data quality and cost are important issues for MCS. The success of the DeepSeek-R1 model released on January 20, 2025, is attributed to advancements in this area. First, data distillation technology is employed [11], thereby obtaining more refined and high-quality data. Based on high-quality data, the hardware requirements and time for training can be significantly reduced, thus cutting huge costs. Another key issue is Privacy-Preserving (PP), which aims to protect sensitive and private information related to workers from being leaked [12], [13], [14]. Private personal information, such as reputation and sensed data, is usually correlated with their jobs, education level, health conditions, etc. [15], [16]. If such information is not well protected, workers may be reluctant to provide their data, which will impede the development of MCS [17], [18]. Despite numerous studies on improving data quality [19], [20], [21] and privacy preserving [10]-[18], there is still room for further research.

Firstly, MCS mainly consists of three parts [22], [23]: Data Requester (DR), Service Platform (SP), and workers [14], [18], where DR submits the scope and quality requirements of sensing data to SP. SP publishes tasks, and workers submit their willingness to perform tasks to SP [14]. SP recruits workers, who sense and submit data to SP. SP calculates the Estimated True Value (ETV) and provides it to DR [14], [25]. Data quality typically refers to the error between the data reported by workers to SP or the ETV provided by SP to DR and the Ground Truth Data (GTD) [20], [21]. Obviously, the smaller the error between the data and GTD, the higher the data quality, thereby enabling the construction of high-quality services. However, obtaining high-quality data poses significant challenges in MCS.

Initially, researchers assumed that the SP had prior knowledge of the quality of the workers and thus selecting high-quality workers could yield high-quality data [26]. Subsequent studies argued that the SP cannot know the quality of workers in advance [19], [27], [28]. However, once the SP receives data from workers, it can determine data quality and thereby evaluate worker quality. This information can then guide the recruitment of high-quality workers to obtain high-quality data [19], [27], [28]. Nevertheless, due to the information elicitation without verification (IEWV) problem in MCS [29], even after receiving data, the SP still cannot ascertain its quality. This renders the aforementioned approaches impractical in real-world MCS applications. To address this, TD methods have been proposed to obtain an ETV close to the GTD [13]-[15], [30]. Among these, the Conflict Resolution on Heterogeneous (CRH) method is one of the most typical approaches [24]. The CRH method

Manuscript received April 21, 2025; revised ***, 2025. This work was supported in part by the Joint Funds of the National Natural Science Foundation of China under Grant U24A20248. (*Corresponding author: Anfeng Liu).

L. Wu, W. Xie, W. Tan and A. Liu are with School of Computer Science and Engineering, Central South University, Changsha 410083 China. (e-mail: 8208221106@csu.edu.cn, xieweikun@csu.edu.cn, wei.tan@csu.edu.cn, afengliu@mail.csu.edu.cn).

T. Wang is with Department of Artificial Intelligence and Future Networks, Beijing Normal University & UIC, Zhuhai, Guangdong, China. (e-mail: tianwang@bnu.edu.cn).

H. Song is with the Department of Information Systems, University of Maryland, Baltimore County (UMBC), Baltimore, MD 21250 USA. (e-mail: h.song@ieee.org).

involves recruiting multiple workers to perform the same task and obtaining the ETV through weighted aggregation of their submissions [24]. In the CRH method, if a worker's submitted data is close to the ETV, their weight is higher; otherwise, it is lower. The CRH method assumes that most workers in MCS are trustworthy [24]. Consequently, the ETV obtained via CRH tends to align with the data submitted by the majority of trustworthy workers, whose weights are higher. This allows the CRH method to achieve a certain level of accuracy in obtaining the ETV without identifying worker quality or trustworthiness. However, the data quality obtained through the CRH method is difficult to guarantee. In the CRH method [24], the presence of even a few dishonest or malicious workers among the recruited workforce can degrade data quality. Moreover, the more untrustworthy workers there are, the greater the quality decline. If multiple malicious workers (MWs) collude in an attack, they can manipulate the SP to obtain values set by the attackers. In MCS, since workers need to move to the task area to sense data, which consumes time, effort, and communication resources, dishonest or malicious workers may submit low-quality, false, or malicious data to minimize or even eliminate costs in order to maximize their profits. Furthermore, due to the IEWV problem in MCS [29], it is challenging to identify the submitted data, which facilitates the reporting of low-quality data and makes it difficult for CRH to ensure data quality.

Privacy-Preserving is another key issue for MCS [10]-[18]. To ensure data quality, privacy preserving is often combined with TD methods, aiming to achieve the dual goals of ensuring data quality and protecting privacy. The privacy-preserving TD method proposed by Tang et al. [14] is a typical example of this category. There are also TD methods based on secret sharing privacy preserving [31] and TD methods based on homomorphic encryption [32]. The common characteristic of these methods is that data acquisition is performed using the CRH method to ensure data quality, and privacy preserving methods are integrated such that the data reported by workers is not leaked to any third party, and even the DR cannot know the data reported by workers [14]. However, as mentioned earlier, current CRH-based methods struggle to ensure data quality, leading to poor performance of these CRH-based privacy-preserving strategies in practical MCS applications.

The reason CRH-based methods cannot ensure data quality is that the calculation of ETV is only related to the data of the current round. Weights are assigned based on the distance between the data reported by workers and the ETV, making it easy for dishonest or malicious workers to report false or malicious data, thereby affecting data quality without punishment. In contrast, reputation-based worker selection strategies select workers based on reputation scores formed by the quality of data reported by workers in the past [33], [34], [35], which can avoid the aforementioned shortcomings. However, the current key issue is the lack of effective methods for obtaining worker reputation scores and integrating these scores into TD methods, which are the issues addressed in this paper.

Although there is no effective method for obtaining worker reputation, researchers have recognized that worker reputation is an important aspect of privacy that needs protection. Wan et al. [2] proposed an approach for protecting worker reputation privacy. Their study did not address how to obtain worker reputation, as it was not the focus of their paper. After worker selection, the data is directly submitted to the DR without passing through the SP, thereby achieving privacy protection of data from the SP. However, this approach does not protect the data privacy from the DR.

Summarizing previous research, the acquisition of high-quality data with privacy protection faces the following challenges: (1) There is a need for a novel TD method that integrates worker reputation to overcome the limitations of traditional CRH methods and obtain high-quality data; (2) An effective method for obtaining worker reputation is required to support reputation-based TD methods; (3) A method with strong privacy protection is needed to provide comprehensive privacy protection for reputation-based TD methods. This involves protecting both reputation privacy and data privacy, ensuring that sensed data remains confidential to workers and is not accessible to other parties, including the DR. (4) The above methods need to be organized into a coordinated system to select high-reputation workers for task execution, obtain high-quality data, and provide comprehensive privacy protection.

To address the above challenges, a Reputation and Data Privacy-Preserving based Truth Discovery (RDPP-TD) scheme is proposed to obtain high-quality data and provide comprehensive privacy protection for MCS. The main innovations of this paper are as follows: (1) We propose a novel Reputation-based Truth Discovery (RTD) method that can obtain worker reputation and high-quality data. In the proposed RTD method, the ETV is no longer determined solely by the weight calculated in the current round, but is also related to the reputation obtained from workers' historical task completion. This approach considers both the data submitted by workers in the current round and their reputation, reducing the impact of workers with low reputation on data quality and enabling the RTD method to effectively improve data quality.

(2) A Reputation and Data Privacy-Preserving (RDPP) approach is proposed to provide privacy preservation for sensed data and reputation values. Firstly, the RDPP approach has carefully designed relevant mechanisms to coordinate with the RTD approach, enabling the system to operate in a privacy-preserving environment. This allows for effective evaluation of worker reputation and obtaining high-quality data while achieving privacy protection for worker reputation and data privacy. Secondly, the RDPP approach also supports reputation-based worker recruitment and rewards, ensuring high-quality data collection and motivating workers to provide accurate data.

(3) Comprehensive theoretical analysis and extensive experiments based on real-world datasets demonstrate that the proposed RDPP-TD scheme provides strong privacy protection and improves data quality by up to 33.3% over previous schemes.

The remainder of this paper is organized as follows. In Section II, we review the existing works. In Section III, the system model and problem formulation are presented. The RDPP-TD scheme is proposed in detail in Section IV. Performance analysis is provided in Section V. Eventually, we draw conclusions and outline future work in Section VI.

Table I. A summary of Privacy-Preserving and TD method of MCS.

| Reference | Data privacy Consideration | Reputation privacy Consideration | TD Consideration | TD method |
|---|---|---|---|---|
| Wan et al. [2] | Matrix encryption | Yes | No | No |
| Tang et al. [14] | Perturbation factor encryption | No | Yes | CRH |
| Feng et al. [31] | Secret-sharing | No | Yes | CRH |
| Zhao et al. [32] | Homomorphic encryption | No | Yes | CRH |
| Wu et al. [33] | No | Yes | Yes | No mentioned |
| RDPP-TD | Perturbation factor encryption | Yes | Yes | Reputation-based TD |

## II. Related Work

MCS is a large-scale, relatively low-cost data collection paradigm, which is of great significance for intelligent big data models [11]. With the rapid advancement of microprocessor technology, the performance of sensing devices has increased several-fold [36], while their size and cost have decreased dramatically. There are currently over 20 billion such sensing devices [37], [38], which are widely distributed. These devices employ participatory sensing, where workers use their portable sensing devices to collect and report data to the SP when they are in or near the task area [39], [40]. High data quality [1], [8], [19]-[21] and privacy preserving [1]-[7], [10]-[18] are two critical issues in MCS data collection. In this section, we present research works related to our study, which mainly focus on data collection strategies based on TD and privacy-preserving strategies for TD-based methods.

*(1) Quality-aware data collection:* it refers to how a SP selects workers to ensure high data quality [1], [8], [19]-[21]. Research in this area has evolved through several phases: (a) Early research assumed the SP could know worker quality in advance [26]. Worker quality was measured by the data they submitted. Thus, obtaining high-quality data meant selecting high-quality workers. However, in MCS, besides data quality, other metrics like cost, delay, and coverage matter [26]. This turns worker selection into a multi-objective optimization problem, which is tough due to conflicting goals [26]. For instance, data quality often correlates linearly with worker cost. So, high-quality data usually means high bids [26]. To handle this, some studies suggested using the ratio of data quality to bid as the worker selection metric, favoring workers with high quality and low bids. Coverage involves both data quality and bids. Liu et al. found that in MCS, worker and task distributions are often uneven and mismatched [26]. This makes it easy to recruit low-bid workers for tasks in crowded areas, while remote tasks, despite the same rewards, struggle to attract workers due to the needed long travel, leading to low coverage. Liu et al. proposed a differentiated rewards method to optimize coverage while keeping costs low. It lowers rewards for hotspot tasks to save costs and raises rewards for remote tasks to encourage worker participation, improving overall task completion quality within a fixed budget [40]. In practical MCS, the SP can't know worker quality in advance, so later studies dropped this assumption.

(b) Subsequent research posited that while the SP doesn't initially know worker quality, it can evaluate it once worker data is received [19], [27], [28]. Based on this, workers can be selected for their quality to obtain high-quality data. Multi-Armed Bandit (MAB)-based worker recruitment is a widely adopted approach [19], [27], [28]. The core idea is that the SP starts with no knowledge of worker quality and uses exploration (trying unknown workers) and exploitation (using known high-quality workers) to select workers. Exploitation means the SP mostly picks from the pool of workers with known good quality, ensuring high data quality [19], [27], [28]. Yet, since this pool is small, the results are only locally optimal. Hence, the MAB strategy also employs exploration, occasionally picking workers of unknown quality, to expand the SP' s knowledge and approach global optimality [19], [27], [28]. In MAB, the Upper Confidence Bound (UCB) index determines worker selection probability [19], [27], [28]. The UCB comprises two parts: worker quality (higher quality means higher priority) and selection frequency (frequently selected workers have a lower chance of being picked again), enabling the SP to explore unknown workers and move toward global optimization [19], [27], [28].

However, in MCS, obtaining worker data quality is challenging [1], [8], [19]-[21], due to the information elicitation without verification (IEWV) problem [29]. IEWV arises because the SP struggles to get GTD, making it hard to assess the quality of workers' reported data. Even after receiving worker data, the SP can't determine its quality or the worker' s reliability. This makes prior studies hard to apply in practice. To address this, TD methods have emerged [3], [4], [13]-[15], [30]. TD methods avoid assessing worker quality and instead recruit multiple workers for the same task [3], [4], [13]-[15], [30]. After collecting their data, the SP calculates the ETV using methods like median [30], mean [41], or weighted mean [24]. The CRH method, which uses a weighted m, initially sets the ETV as the average of the collected data. It then computes each data point' s distance to the ETV, sums these distances, and calculates each worker' s weight as their distance divided by the total distance sum, taking the log of this value [24]. The ETV is updated as the weighted sum of workers' s data and weights. This process repeats recomputing differences, weights, and ETV until two consecutive ETVs are within a set threshold, yielding the final ETV.

TD methods like CRH have the edge of broad applicability,

needing no worker quality assessment. Yet, their biggest drawback is no guarantee of data quality. Even a few dishonest or malicious workers can degrade data quality. Worse, if over half the workers are MWs, the ETV becomes whatever they dictate, making CRH entirely ineffective.

*(2) Privacy preserving methods:* PP research relevant to this paper mainly focuses on the privacy protection of worker reputation and sensed data. Firstly, there are methods focusing on the privacy of sensed data, aiming to prevent the SP or any third party from accessing the data submitted by workers [14]. Below, we introduce a category of privacy-preserving methods for sensed data based on TD [14], [31], [32]. The primary characteristic of these methods is modifying the CRH method to incorporate privacy-preserving functionalities [14], [31], [32]. Essentially, these privacy-preserving methods primarily emphasize privacy protection, while data quality remains dependent on the CRH method [14], [31], [32]. As discussed earlier, the CRH method does not sufficiently ensure data quality. Therefore, this paper proposes a novel RTD method with privacy protection for both sensed data and worker reputation, aiming to achieve higher data quality than previous strategies. Current TD-based privacy-preserving research can be categorized as follows: encryption-based [32], [36], secret sharing-based [31], anonymity-based [5], [33], [42], and lightweight methods [14].

(a) Encryption-based privacy preserving is a common approach where workers upload encrypted data. While this protects privacy, TD methods often require computations (e.g., logs, multiplication, division) on the data to compute ETV, which most encryption methods do not support [36]. Homomorphic encryption is a potential solution as it allows computations on encrypted data, yielding results matching those from computations on unencrypted data [36]. However, homomorphic encryption often does not support all necessary operations. Zhao et al. [32] proposed a privacy-preserving method using custom protocols (SecLog, SecMulDiv, SecDist) integrated into the TD strategy PRICE [32].

(b) Secret-sharing privacy preserving involves splitting data into shares distributed across servers so that no single server can access the full data. For example, Feng et al. [31] introduced EPRICE, where the SP comprises three non-colluding servers. Workers split their data into three shares, and EPRICE defines privacy-preserving operations (privacy multiplication protocols, privacy logarithm protocols, and privacy division protocols.) for CRH-based TD. The three servers collaboratively perform these operations to achieve privacy protection while computing the ETV [31].

(c) Anonymity-based privacy preserving is another effective approach. Wu et al. [33] proposed an anonymous authentication protocol using random pseudonyms and reputation privacy protection. The randomness of pseudonyms breaks the link between a user' s real identity and their pseudonym, ensuring privacy protection [33].

(d) Lightweight TD-based privacy preserving methods. Tang et al. [14] proposed a lightweight privacy-preserving TD method. Their method introduces an Assistant Server (AS) to work with the DR, enabling privacy-preserving truth discovery based on the CRH method [24].

While TD-based privacy-preserving methods are widely applicable, they inherit data quality limitations of CRH. Reputation-based methods address this by selecting high-quality workers. Below, we discuss privacy-preserving methods integrating reputation mechanisms.

*(3) Reputation-based privacy preserving methods:* Research on reputation-based privacy preserving methods [2], [10], [12] is still in development. Current studies primarily focus on protecting and accessing reputation information [16], [33]-[35]. For instance, Gao et al. [16] addressed the challenge of comparing and selecting workers with high reputation across different domains without revealing their actual reputation values. Wan et al. [2] proposed a method where worker reputation is managed by a Trust Authority (TA), allowing the selection of workers based on reputation thresholds while preserving privacy and allocating rewards accordingly. Wu et al. [33] leveraged blockchain technology, storing reputation and sensed data separately in a reputation chain and a sensing chain to enable anonymous access to reputation information.

Compared to existing reputation-based privacy-preserving methods, the method proposed in this paper addresses two key issues. Firstly, it tackles the evaluation and acquisition of worker reputation. Current research often overlooks how reputation is obtained, with some studies assuming it is system-inherent [16] and others relying on DR feedback [33]. However, establishing a clear method for obtaining reputation is crucial for MCS. Second, it effectively integrates reputation into TD to enhance data quality. Unlike previous work that focuses mainly on protecting and accessing reputation, our method emphasizes using reputation to guide worker selection and improve the quality of collected data.

## III. System model and Problem Formulation

### A. System model

The MCS system primarily consists of a service platform, data requesters, workers, and a trust authority. As depicted in Fig. 1, the service platform strategically recruits workers who are equipped with mobile sensor devices to meet the requesters' needs for on-requirement sensory data and to generate predictive outcomes. Concurrently, the trust authority ensures the privacy protection of the system.

**Definition 1 (Service Platform)** The service platform, equipped with robust storage and computing capabilities, facilitates task distribution, worker recruitment, and reward allocation.

**Definition 2 (Trust Authority)** The trust authority serves as a fully trusted entity within the MCS system, playing a crucial role in safeguarding the privacy and security of system.

**Definition 3 (Data Requester, Tasks)** The data requester is an enterprise or individual that seeks specific data who publishes a set of $M$ tasks, denoted by $\mathcal{M} = \{m_1, m_2, \cdots, m_M\}$, to the service platform. Data collection occurs in successive rounds, indexed by $t \in \{1,2,3\cdots\}$.

**Definition 4 (Task Requirements)** Each task $m_i$ in round t has its own set of requirements $\mathcal{D}_{i,t} = \{\mathcal{T}_{i,t}, \mathcal{L}_{i,t}, B_{i,t}, C_{i,t}\}$, which specify the detailed requirements for time, longitude, latitude, and reputation, respectively.

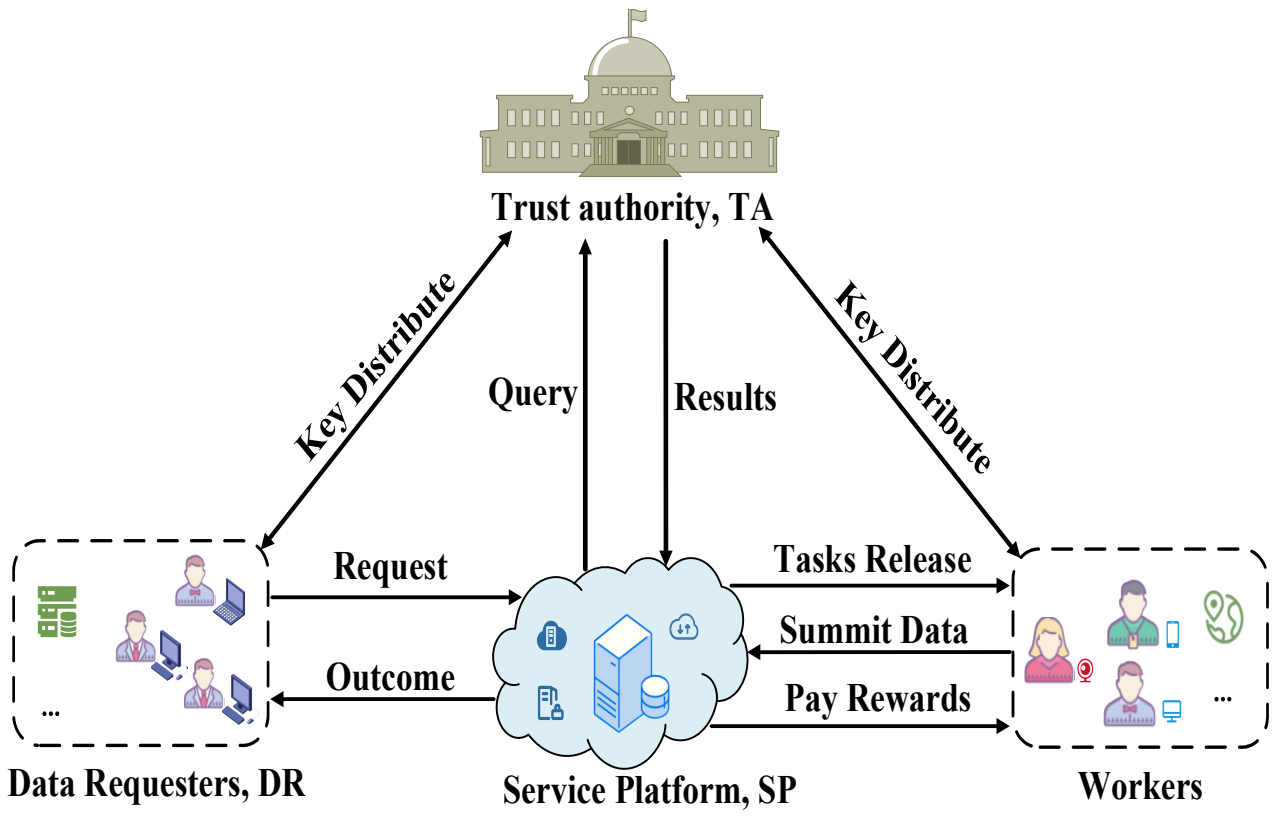

**Fig. 1**. System framework of RDPP-TD

The requirements $\mathcal{D}_{i,t}$ are specified as follows:

(1) Time requirement: The acceptable time frame for the task $m_{i,t}$ is denoted by $\mathcal{T}_{i,t} \in [\overleftarrow{t_{i,t}}, \overrightarrow{t_{i,t}}]$, where $\overleftarrow{t_{i,t}}$ and $\overrightarrow{t_{i,t}}$ represent the start and end times of the time frame.

(2) Longitude requirement: The acceptable longitudinal area for the task $m_{i,t}$ is denoted by $\mathcal{L}_{i,t} \in [\overleftarrow{\ell_{i,t}}, \overrightarrow{\ell_{i,t}}]$, where $\overleftarrow{\ell_{i,t}}$ and $\overrightarrow{\ell_{i,t}}$ represent the minimum and maximum longitudinal boundaries.

(3) Latitude requirement: The acceptable latitudinal area for the task $m_{i,t}$ is denoted by $B_{i,t} \in [\overleftarrow{b_{i,t}}, \overrightarrow{b_{i,t}}]$, where $\overleftarrow{b_{i,t}}$ and $\overrightarrow{b_{i,t}}$ represent the minimum and maximum latitudinal boundaries.

(4) Reputation requirement: The reputation requirement for the task $m_{i,t}$ is denoted by $C_{i,t} = c_{i,t}$, where $c_{i,t}$ represent the threshold value of reputation.

**Definition 5 (Workers)** Consider a set of $N$ workers $S = \{s_1, s_2, \cdots, s_N\}$. Each worker $s_j$ is characterized by a set of attributes $\mathcal{D}_{i,j}^t = \{\mathcal{T}_{i,j}^t, \mathcal{L}_{i,j}^t, B_{i,j}^t, C_{j,t}\}$. Attributes $\mathcal{D}_{i,j}^t$ serves as the basis for evaluating worker $s_j$'s suitability for participating in task $m_{i,t}$.

The attributes $\mathcal{D}_{i,j}^t$ are specified as follows:

(1) Time Attribute: $\mathcal{T}_{i,j}^t$ represents the worker $s_j$'s current time status for task $m_{i,t}$.

(2) Longitude Attribute: $\mathcal{L}_{i,j}^t$ specifies the geographical longitude of worker $s_j$ for task $m_{i,t}$.

(3) Latitude Attribute: $B_{i,j}^t$ specifies the geographical latitude of worker $s_j$ for task $m_{i,t}$.

(4) Reputation Attribute: $C_{j,t}$ represents the reputation of worker $s_j$ for task $m_{i,t}$.

**Definition 6 (Recruitment Results).** Worker $s_j$ can accept a task $m_{i,t}$ only if attributes $\mathcal{T}_{i,j}^t, \mathcal{L}_{i,j}^t, B_{i,j}^t$ *and* $C_{j,t}$ meet the requirements $\mathcal{D}_{i,t}$ of the task. After evaluating whether the worker $s_j$ meets the requirements $\mathcal{D}_{i,t}$, the SP sends a binary variable $r_{i,j}^t$ to the worker $s_j$. Specifically, $r_{i,j}^t = 1$ if worker $s_j$ is recruited for task $m_{i,t}$, and $r_{i,j}^t = 0$ otherwise.

**Definition 7 (Sensed Data, Ground Truth).** The sensed data, collected via mobile devices by workers, are denoted by the set $\mathcal{X}_{i,t} = \{x_{i,1}^t, x_{i,2}^t, \cdots, x_{i,n}^t\}$, where each $x_{i,j}^t$ represents the data contributed by worker $s_j$ for task $m_{i,t}$. The ground truth is real value for task $m_{i,t}$, which is denoted by $g_{i,t}$.

**Definition 8 (Estimated Truth).** The estimated truth for task $m_{i,t}$ is denoted by $\tilde{x}_{i,t}$, which is computed from the sensed data $\mathcal{X}_{i,t}$. The estimated truth $\tilde{x}_{i,t}$ is an approximation for ground truth $g_{i,t}$, offering a solution while the ground truth is difficult to access.

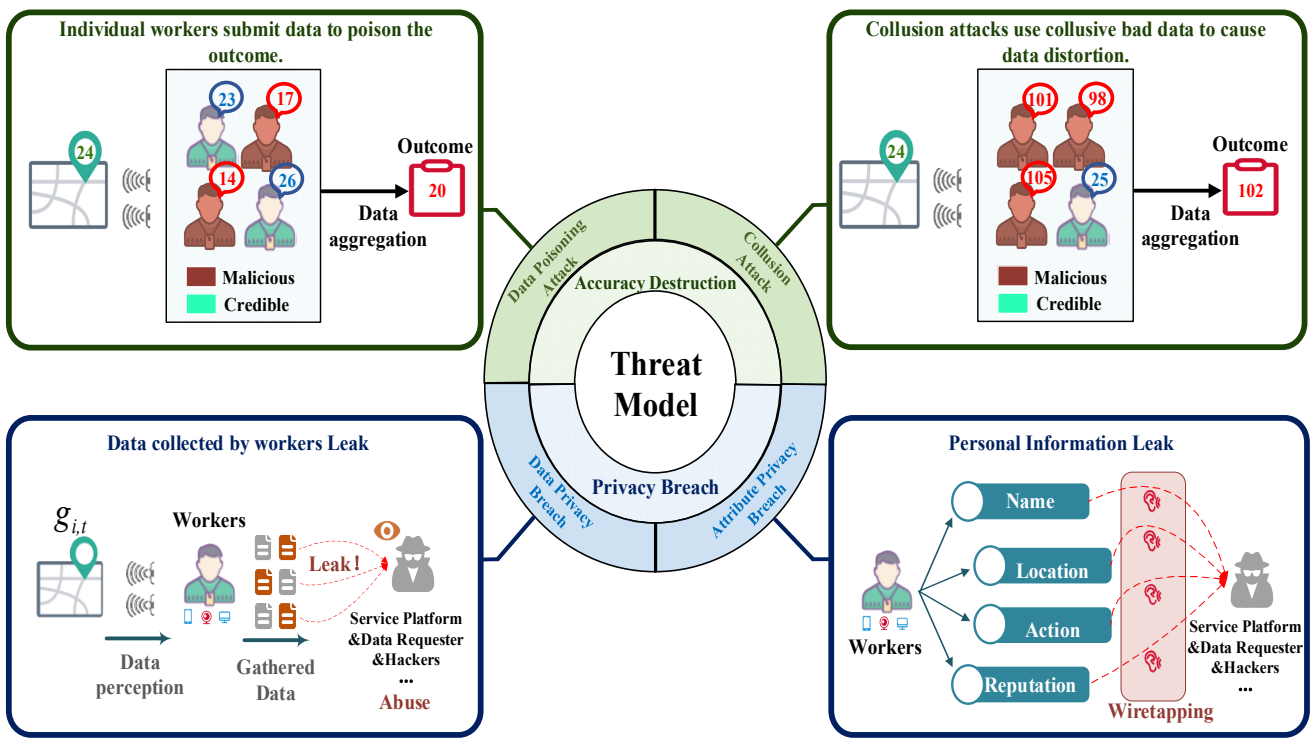

**Fig. 2**. Threat Model.

## *B. Threat model*

As shown in Fig. 2, the MCS system faces several potential threats to its integrity and security.

- **Data Poisoning Attack:** Some workers may intentionally upload incorrect data or submit unreliable data that fails to meet the task requirements, thereby undermining the accuracy of the task outcomes.
- **Collusion Attack:** Certain workers may collude by submitting similar data, which can mislead the computation of task results and enable them to fraudulently claim task rewards.
- **Attribute Privacy Breach:** Workers' sensitive attribute information, including their names, location, action and reputation, could be vulnerable to breaches, potentially harming their privacy and security.
- **Data Privacy Breach:** Data collected by workers may be at risk of being misused by the service platform, data requesters, or other entities.

## *C. Problem Formulation*

This paper introduces an innovative scheme RDPP-TD for MCS, with the overarching goal of achieving efficient worker recruitment and truth discovery while ensuring comprehensive privacy protection for workers. The optimization objectives are given as follows:

(1) Minimize the RMSE of ETV.

Estimate data $\tilde{x}_{i,t}$ refers to ETV aggregated from sensed data set $\mathcal{X}_{i,t}$, with the goal of approximating GTD as closely as possible. We use $\Re$ to evaluate the accuracy of the data in task $m_{i,t}$ based on Root Mean Square Error (RMSE) method. The data accuracy is calculated as Eq. (1), where a smaller $\Re$ correlates with greater accuracy.

***Minimize.***

$$\mathfrak{R} = \sqrt{\frac{1}{m}\sum_{k=1}^{m}\left(\tilde{x}_{i,t} - g_{i,t}\right)^2} \quad (1)$$

***Subject to.***

$$\forall \psi_{i,j}^t = 1, \mathcal{D}_{i,j}^t \text{ satisfy } \mathcal{D}_{i,t} \quad (2)$$

Our first optimization objective is to minimize the value of $\mathfrak{R}$ and improve data service accuracy, while fulfilling the constraint specified in Eq. (2).

(2) Maximize data quality of ETV.

The data quality is an indicator of the accuracy of TD method. Therefore, we use $\mathcal{Q}_{i,t}$ to evaluate the data quality of the estimate data $\tilde{x}_{i,t}$ for task $m_{i,t}$, which is calculated as Eq. (3) and Eq. (4).

$$\mathcal{Q}_{i,t} = 1 - \left|\left(\frac{1}{1 + \lambda^{-q_{i,t}}} - 0.5\right) * 2\right|. \quad (3)$$

$$q_{i,t} = \frac{\left|\tilde{x}_{i,t} - g_{i,t}\right|}{\tilde{x}_{i,t}}. \quad (4)$$

Where $\mathcal{Q}_{i,t} \in [0,1]$ and $\lambda > 1$. $\lambda$ is the base of the exponent. Our second optimization objective is to maximize the average value of $\mathcal{Q}_{i,t}$ as Eq. (5), while fulfilling the constraint specified in Eq. (6).

***Maximize.***

$$\mathcal{Q} = \frac{1}{m}\sum_{k=1}^{m}\mathcal{Q}_{i,t}. \quad (5)$$

***Subject to.***

$$\forall \psi_{i,j}^t = 1, \mathcal{D}_{i,j}^t \text{ satisfy } \mathcal{D}_{i,t} \quad (6)$$

(3) Minimize the MAE of reputation value.

True reputation $c_{j,t}$ refers to probability that the worker $s_j$ submits high quality data in round $t$. Reputation value $C_{j,t}$ is an estimate of true reputation $c_{j,t}$, which is denoted as Eq. (7).

$$C_{j,t} = \frac{\alpha_{j,t}}{\alpha_{j,t} + \beta_{j,t}}. \quad (7)$$

Here, $\alpha_{j,t}$ represents the count of successful tasks and $\beta_{j,t}$ represents the count of failed tasks for the worker $s_j$ up to round $t$. We use $\mathfrak{P}$ to evaluate the accuracy of reputation value $C_{j,t}$ based on Mean Absolute Error (MAE) method as Eq. (8).

***Minimize.***

$$\mathfrak{P} = \frac{1}{n}\sum_{k=1}^{n}\left|C_{j,t} - c_{j,t}\right| \quad (8)$$

Our optimization objective is to minimize the value of $\mathfrak{P}$ and achieve accurate identification of worker reputation.

(4) Maximize the robustness under collusion attacks.

Collusion attacks in MCS system refer to scenarios where malicious workers (MWs) conspire to submit similar false data to mislead ETV. The robustness to resist collusion attacks is significant in MCS system, therefore, our fourth optimization objective is to minimize RMSE and maximize data quality as Eq. (1) and Eq. (5) under collusion attacks.

To enhance the clarity of the article, Table II is provided to represent all the common notations used in this paper.

Table II. Description of major notations.

| Notations | Description |
|---|---|
| $\mathcal{M}$ | Set of tasks |
| $S$ | Set of workers |
| $m_{i,t}$ | Task $m_i$ in round t |
| $s_j$ | Worker $j$ |
| $\mathcal{D}_{i,t}$ | Requirement for task $m_{i,t}$ |
| $\mathcal{T}_{i,t}$ | Time requirement for task $m_{i,t}$ |
| $\mathcal{L}_{i,t}$, $B_{i,t}$ | Longitude, Latitude requirement for task $m_{i,t}$ |
| $C_{i,t}$ | Reputation requirement for task $m_{i,t}$ |
| $\mathcal{D}_{i,j}^t$ | Attributes of $s_j$ in task $m_{i,t}$ |
| $\mathcal{T}_{i,j}^t$ | Time of $s_j$ in task $m_{i,t}$ |
| $\mathcal{L}_{i,j}^t$, $B_{i,j}^t$ | Longitude, Latitude of $s_j$ in task $m_{i,t}$ |
| $C_{j,t}$ | Reputation value of $s_j$ in task $m_{i,t}$ |
| $x_{i,j}^t$ | Data collected by $s_j$ in task $m_{i,t}$ |
| $g_{i,t}$, $\tilde{x}_{i,t}$ | Ground truth, estimate truth in task $m_{i,t}$ |
| $y_{i,j}^t$, $z_{i,j}^t$ | Additive, multiplicative noise data from $s_j$ for task $m_{i,t}$ |
| $w_{i,j}^t$ | Weight of $s_j$ for task $m_{i,t}$ |
| $\psi_{i,j}^t$ | Recruitment result for $s_j$ in task $m_{i,t}$ |
| $\mathfrak{R}$ | RMSE of ETV |
| $\mathcal{Q}_{i,t}$ | Data Quality of ETV |
| $\mathfrak{P}$ | Worker identification accuracy |

## IV. The design of the RDPP-TD scheme

### *A. The overview of RDPP-TD scheme*

The traditional MCS systems are confronted with numerous challenges. The privacy information of workers and the collected data are at risk of being eavesdropped on and misused by malicious attackers. Moreover, the existing CRH data aggregation method lacks robustness when facing malicious attacks. To address these issues, this paper proposes the RDPP-TD scheme. The RDPP-TD scheme employs a method based on matrix encryption and perturbation factors to achieve comprehensive privacy protection for workers. It also introduces an improved method for CRH data aggregation, namely the RTD method, which enhances the accuracy and the robustness. Additionally, the RDPP-TD scheme integrates the Elgamal public-key cryptosystem to ensure the security of communications. As illustrated in Fig.3, the specific workflow of the RDPP-TD scheme is elaborated as follows:

(1) System Initialization: the trust authority (TA) establishes the foundational cryptographic components by selecting a prime constant $p$ and forming a cyclic group $C$ of order $p$, with $c$ serving as its generator. The TA generates a public/private key pair, ($pk_s$, $sk_s$), where $sk_s$ is randomly drawn from $\mathbb{Z}_p^*$ and $pk_s$ = ($c$, $h$, $p$) with $c, h, p \in \mathbb{Z}_p^*$.

(2) The registration of the workers: each worker $s_j$ is assigned a uniquely pseudonym $a_j$ by TA.

$$a_j = n_j \oplus \mathcal{H}(sk_s \cdot salt). \quad (9)$$

$n_j$ represents the real identity of the worker $s_j$. Salt is a randomly selected value from $\mathbb{Z}_p^*$ and is stored securely in the TA's database. The function $\mathcal{H}(x)$ represents a cryptographic Hash function, and $\oplus$ denotes the exclusive OR operation.

(2) Task publishing: the data requester (DR), aiming to initiate a sensing task $m_{i,t}$, generates a public/private key pair $(pk_r, sk_r)$ and registers with the TA. The task $m_{i,t}$, along with the budget $\mathfrak{B}_{i,t}$ and the range reliability requirement $\mathcal{D}_{i,t}$ is communicated to the service platform (SP), who then releases the task $m_{i,t}\{\mathcal{D}_{i,t}, \mathfrak{B}_{i,t}\}$ to the potential workers.

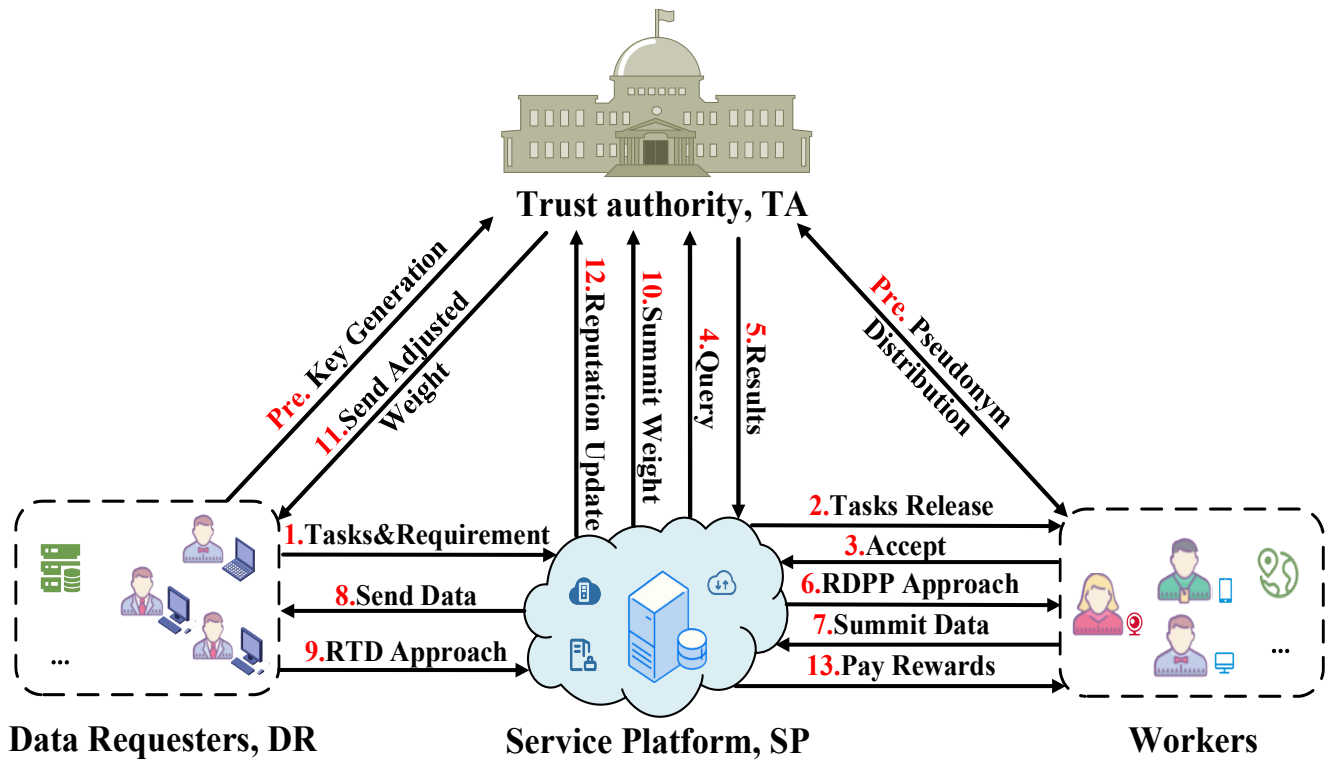

**Fig. 3**. The overview of RDPP-TD scheme.

(4) Worker engagement: After receiving the task $m_{i,t}$ and deciding to participate, worker $s_j$ generates current time $\mathcal{T}_{i,j}^t$, longitude $\mathcal{L}_{i,j}^t$, latitude $B_{i,j}^t$, and makes an initial commitment $\mathcal{D}_{i,j}^t$ to the SP. The calculation of the commitment is detailed in section IV-B.

(5) Secure worker recruitment: upon receiving the worker $s_j$'s commitment $\mathcal{D}_{i,j}^t$, SP queries TA whether the reputation value $C_{j,t}$ of worker $s_j$ is greater than or equals to the reputation threshold $C_{i,t}$. If worker $s_j$ meets the reputation requirement, SP then selects qualified workers through the RDPP approach. The details of RDPP approach are explained in section IV-B.

(6) Data collection and submission: after RDPP process, qualified workers collect data $x_{i,j}^t$ and encrypt the data to obtain $y_{i,j}^t$, $z_{i,j}^t$. The encryption process is detailed as follows:

$$\begin{cases} y_{i,j}^t = x_{i,j}^t + \alpha_{i,j}^t \\ z_{i,j}^t = x_{i,j}^t \cdot \beta_{i,j}^t \end{cases} \tag{10}$$

Subsequently, workers encrypt $y_{i,j}^t$ and $z_{i,j}^t$ using the public key $pk_r$ of DR and transmit the encrypted data to SP.

$$Enc((y_{i,j}^t, z_{i,j}^t), pk_r) \tag{11}$$

(7) Data aggregation: after receiving the encrypted data, SP send it to DR and start the RTD-based data aggregation process. DR calculates the distance from worker's data to estimated true in $n^{th}$ iteration and send it to SP.

$$d(y_{i,j}^t, \tilde{x}_{i,t}^n) = y_{i,j}^t - \tilde{x}_{i,t}^n \tag{12}$$

SP then calculates the weight $w_{i,j}^t$ of worker by adjusting the $d(y_{i,j}^t, \tilde{x}_{i,t}^n)$ and sends it to TA. TA calculate the adjusted weight $\widetilde{w}_{i,j}^t$ and the sum of adjusted weight $\sigma_{w_{i,j}^t}$'. Then, TA send $\widetilde{w}_{i,j}^t$ and $\sigma_{w_{i,j}^t}$' to DR, which enables DR to calculate the $(n+1)^{th}$ estimated true. The detail calculation is explained in the section IV-C.

(8) Rewards allocation: after the final iteration, SP allocates rewards to each worker based on the quality of their submitted data. $\mathcal{R}_{i,j}^t$ represents the reward for $s_j$ in task $m_{i,t}$ and is calculated according to Eq. (13).

$$\mathcal{R}_{i,j}^t = \left(\frac{w_{i,j}^t}{\sum_{j=1}^k w_{i,j}^t}\right) \cdot \mathfrak{B}_{i,t}. \tag{13}$$

(9) Reputation update: SP compares the distance $d(x_{i,j}^t, \tilde{x}_{i,t}^n)$ with parameter $\gamma$. If $d(x_{i,j}^t, \tilde{x}_{i,t}^n)$ is less than or equals to $\gamma$, $(a_j, 1)$ is sent to TA, otherwise $(a_j, 0)$ is sent. Then, TA update the reputation of workers as follows:

$$\alpha_{j,t+1}, = \begin{cases} \alpha_{j,t} + 1, & if(a_j, 1) \\ \alpha_{j,t}, & else \end{cases} \tag{14}$$

$$\beta_{j,t+1}, = \begin{cases} \beta_{j,t} + 1, & if(a_j, 0) \\ \beta_{j,t}, & else \end{cases} \tag{15}$$

RDPP-TD incentivizes workers by offering increased rewards for superior data quality and enhances their competitiveness through improved reputation, which in turn motivates workers to submit high-quality data.

## *B. RDPP-based Worker Recruitment*

The quality of recruited workers and the match between workers and task requirements are crucial for the computation of the ETV in MCS. Therefore, this paper proposes an RDPP-based worker recruitment approach that can efficiently screen workers while protecting their privacy. The detail process is as follows:

After receiving the task and deciding to participate, worker $s_j$ are required to generate three attribute vectors $v_{\mathcal{T}_{i,j}^t}$, $v_{\mathcal{L}_{i,j}^t}$, and $v_{B_{i,j}^t}$. These vectors are combined into $v_{\mathcal{D}_{i,j}^t}$, and the details of generation are as follows:

$$v_{\mathcal{D}_{i,j}^t} = \begin{cases} v_{\mathcal{T}_{i,j}^t} = ({\mathcal{T}_{i,j}^t}^3, {\mathcal{T}_{i,j}^t}^2, \mathcal{T}_{i,j}^t, 1) \\ v_{\mathcal{L}_{i,j}^t} = ({\mathcal{L}_{i,j}^t}^3, {\mathcal{L}_{i,j}^t}^2, \mathcal{L}_{i,j}^t, 1) \\ v_{B_{i,j}^t} = ({B_{i,j}^t}^3, {B_{i,j}^t}^2, B_{i,j}^t, 1) \end{cases} . \tag{16}$$

Each of these vectors is then encrypted to ensure data privacy. For simplicity, the encryption process for $v_{\mathcal{D}_{i,j}^t}$ will be described, which applies to each of the three vectors individually. The encrypted vector $E_{\mathcal{K}}(v_{\mathcal{D}_{i,j}^t})$ is computed and submitted to SP. The encryption is performed as follows:

$$E_{\mathcal{K}}(v_{\mathcal{D}_{i,j}^t}) = \tau_{i,j}^t \cdot |det(\mathcal{K})| \cdot \mathcal{K}^{-1} \cdot v_{D_{i,j}^t} \tag{17}$$

$\tau_{i,j}^t$ is a randomly generated $\mu$-bit positive integer. A cryptographic key $\mathcal{K}$ is defined for encrypting 4-dimensional vector in a $4 \times 4$ matrix format, and $\mathcal{K}^{-1}$ represents the inverse of matrix $\mathcal{K}$. $E_{\mathcal{K}}(v_{\mathcal{D}_{i,j}^t})$ acts as a commitment, protecting worker privacy and allowing SP to verify the eligibility of the worker against the range reliability requirement.

Upon receiving the worker $s_j$'s commitment $E_{\mathcal{K}}(v_{\mathcal{D}_{i,j}^t})$, SP

sends queries $(a_j, C_{i,t})$ to TA to judge whether $s_j$ meets the reputation requirement. TA return the results $r_{i,j}^t$ according to Eq. (18):

$$r_{i,j}^t = \begin{cases} 0, & if\ C_{j,t} \geq C_{i,t} \\ 1, & else \end{cases} \tag{18}$$

If $r_{i,j}^t = 1$, SP then dispatches the set of conditions $\mathcal{D}_{i,t}$ necessary for $m_{i,t}$ to the qualified worker. The generation of $\mathcal{D}_{i,t}$ is as follows:

$$\mathcal{D}_{i,t} = \begin{cases} \mathcal{T}_{i,t} = \left[\overleftarrow{t_{i,t}}, \overrightarrow{t_{i,t}}\right] \\ \mathcal{L}_{i,t} = \left[\overleftarrow{\ell_{i,t}}, \overrightarrow{\ell_{i,t}}\right] \\ \mathcal{B}_{i,t} = \left[\overleftarrow{b_{i,t}}, \overrightarrow{b_{i,t}}\right] \end{cases} \tag{19}$$

After receiving the $\mathcal{D}_{i,t}$, workers evaluate whether they satisfy the requirements. If they comply with the requirements $\mathcal{D}_{i,t}$, they then proceed to calculate the verification vector set $d_{i,j}^t = \{d_{\mathcal{T}_{i,j}^t}, d_{\mathcal{L}_{i,j}^t}, d_{\mathcal{B}_{i,j}^t}\}$. The detailed process for computing $d_{i,j}^t$ is outlined below:

$$d_{i,j}^t = \begin{cases} (1, -\overleftarrow{t_{i,t}} - \overrightarrow{t_{i,t}} + \varepsilon, \overleftarrow{t_{i,t}} \cdot \overrightarrow{t_{i,t}} - \overleftarrow{t_{i,t}} \cdot \varepsilon - \overrightarrow{t_{i,t}} \cdot \varepsilon, \overleftarrow{t_{i,t}} \cdot \overrightarrow{t_{i,t}} \cdot \varepsilon)^T \\ (1, -\overleftarrow{\ell_{i,t}} - \overrightarrow{\ell_{i,t}} + \varepsilon, \overleftarrow{\ell_{i,t}} \cdot \overrightarrow{\ell_{i,t}} - \overleftarrow{\ell_{i,t}} \cdot \varepsilon - \overrightarrow{\ell_{i,t}} \cdot \varepsilon, \overleftarrow{\ell_{i,t}} \cdot \overrightarrow{\ell_{i,t}} \cdot \varepsilon)^T \\ (1, -\overleftarrow{b_{i,t}} - \overrightarrow{b_{i,t}} + \varepsilon, \overleftarrow{b_{i,t}} \cdot \overrightarrow{b_{i,t}} - \overleftarrow{b_{i,t}} \cdot \varepsilon - \overrightarrow{b_{i,t}} \cdot \varepsilon, \overleftarrow{b_{i,t}} \cdot \overrightarrow{b_{i,t}} \cdot \varepsilon)^T \end{cases} \tag{20}$$

Subsequently, the vector $d_{i,j}^t$ is scaled by a random number $\tau_{i,j}^{t\,\prime}$ and then multiplied by the transpose of the encryption matrix $\mathcal{K}$, and submitted to SP.

$$E_{\mathcal{K}}(d_{i,j}^t) = \tau_{i,j}^{t\,\prime} \cdot \mathcal{K}^T \cdot d_{i,j}^t \tag{21}$$

Upon receiving $E_{\mathcal{K}}(d_{i,j}^t)$, SP verifies it by performing the $[A \cdot B]$ operation with the initial commitment made by the worker. The verification checks if the result is greater than or equals to zero.

$$\left[E_{\mathcal{K}}(v_{\mathcal{D}_{i,j}^t}) \cdot E_{\mathcal{K}}(d_{i,j}^t)\right] \geq 0 \tag{22}$$

If the verification succeeds, SP select eligible workers and employs stratified sampling algorithm to select the required number for $m_{i,t}$. A pair of perturbation factors $(\alpha_{i,j}^t, \beta_{i,j}^t)$ is shared between SP and the selected worker. Recruitment result $\psi_{i,j}^t \in \{0,1\}$ is sent to workers, where $\psi_{i,j}^t = 1$ represents selected and 0 otherwise.

The RDPP-based worker recruitment is described in **Algorithm 1**.

**Algorithm 1**: RDPP

**Input**: $\mathcal{D}_{i,t}, \mathcal{D}_{i,j}^t$

**Output**: $\psi_{i,j}^t \in \{0,1\}$

```
Broadcast m_{i,t} to workers.
  For each s_j do
      Generate v_{D^t_{i,j}} = (T^t_{i,j}, L^t_{i,j}, B^t_{i,j});
      Encrypt v_{D^t_{i,j}} → E_K(v_{D^t_{i,j}});
    └ Send E_K(v_{D^t_{i,j}}) to SP;
  If C_{j,t} > C_{i,t}
  └ SP send D_{i,t} to qualified workers;
  For each qualified s_j do
      If D^t_{i,j} satisfies D_{i,t}
          Calculate d^t_{i,j};
          Encrypt d^t_{i,j} → E_K(d^t_{i,j});
        └ Send E_K(d^t_{i,j}) to SP;
  If [E_K(v_{D^t_{i,j}}) · E_K(d^t_{i,j})] ≥ 0
  └ SP send ψ^t_{i,j} = 1 to qualified workers;
  Else SP send ψ^t_{i,j} = 0;
└ Return (ψ^t_{i,j})
```

*C. RTD-based Data Aggregation*

In traditional data aggregation method CRH, the computation of ETV is only related to the data of the current round. As a result, the quality is difficult to ensure and cannot withstand collusive attacks launched by MWs. However, the RTD-based data aggregation method considered the reputation formed based on the workers' historical task completion, which can achieve high-quality ETV and enhance the system robustness against collusive attacks. The detail process is as follows:

Once received the encrypted data sent by SP, DR decrypts the message and gets $y_{i,j}^t, z_{i,j}^t$. The first step involves adjusting the received data by subtracting the previous iteration's ETV $\widetilde{x}_{i,t}^n$, which initialized to a non-zero value $\widetilde{x}_{i,t}^0$ according to Eq. (12). $d(y_{i,j}^t, \widetilde{x}_{i,t}^n)$ is then forwarded to SP, who, being aware of perturbation factors $(\alpha_{i,j}^t, \beta_{i,j}^t)$, can deduce the distance from each worker's data to $\widetilde{x}_{i,t}^n$ after accounting for the noise:

$$d(x_{i,j}^t, \widetilde{x}_{i,t}^n) = d(y_{i,j}^t, \widetilde{x}_{i,t}^n) - \alpha_{i,j}^t = x_{i,j}^t - \widetilde{x}_{i,t}^n \tag{23}$$

SP then proceeds to calculate the corresponding weights $w_{i,j}^t$ for each worker's contribution. The calculation of $w_{i,j}^t$ is as Eq. (24). SP send $(w_{i,j}^t, \beta_{i,j}^t)$ to TA, who then calculates adjusted weight $\widetilde{w}_{i,j}^t$ and the sum of adjusted weight $\sigma_{w_{i,j}^t{}'}$ as Eq. (25) and Eq. (26).

$$w_{i,j}^t = log\left(\frac{\sum_{j=1}^{k} d(x_{i,j}^t, \widetilde{x}_{i,t}^n)}{d(x_{i,j}^t, \widetilde{x}_{i,t}^n)}\right) \tag{24}$$

$$\sigma_{w_{i,j}^t{}'} = \sum_{j=1}^{k} (w_{i,j}^t \cdot C_{j,t}) \tag{25}$$

Where, $k$ is number of selected workers.

$$\widetilde{w}_{i,j}^t = \frac{w_{i,j}^t \cdot C_{j,t}}{\beta_{i,j}^t} \tag{26}$$

Upon receiving $\widetilde{w}_{i,j}^t$ and $\sigma_{w_{i,j}^t{}'}$ by TA, DR then multiplies the data $z_{i,j}^t$ by $\widetilde{w}_{i,j}^t$ to obtain a weighted data set and normalizes the weighted data from workers by the sum of the weights multiplied by the reputation to aggregate the ETV for the current $(n+1)^{th}$ iteration as shown in Eq.(27) and Eq.(28).

$$z_{i,j}^t \cdot \widetilde{w}_{i,j}^t = x_{i,j}^t \cdot w_{i,j}^t \cdot C_{j,t} \tag{27}$$

$$\widetilde{x}_{i,t}^{n+1} = \frac{\sum_{j=1}^{k} x_{i,j}^t \cdot w_{i,j}^t \cdot C_{j,t}}{\sigma_{w_{i,j}^t{}'}} \tag{28}$$

The iteration continues until a specific number of rounds $\mathcal{N}$ is reached or the difference between the current $\widetilde{x}_{i,t}^n$ and the previous $\widetilde{x}_{i,t}^{n-1}$ is less than a predefined threshold $\delta$. The RTD-based data aggregation workflow is depicted as Fig. 4.

$$\left|\widetilde{x}_{i,t}^n - \widetilde{x}_{i,t}^{n-1}\right| < \delta \tag{29}$$

## D. Theoretical Analysis

In this section, we conduct a theoretical analysis of the security of the RDPP-TD scheme, focusing on the privacy protection of workers in RDPP approach and the data security of workers in RTD process, as mentioned in sections IV-B and IV-C.

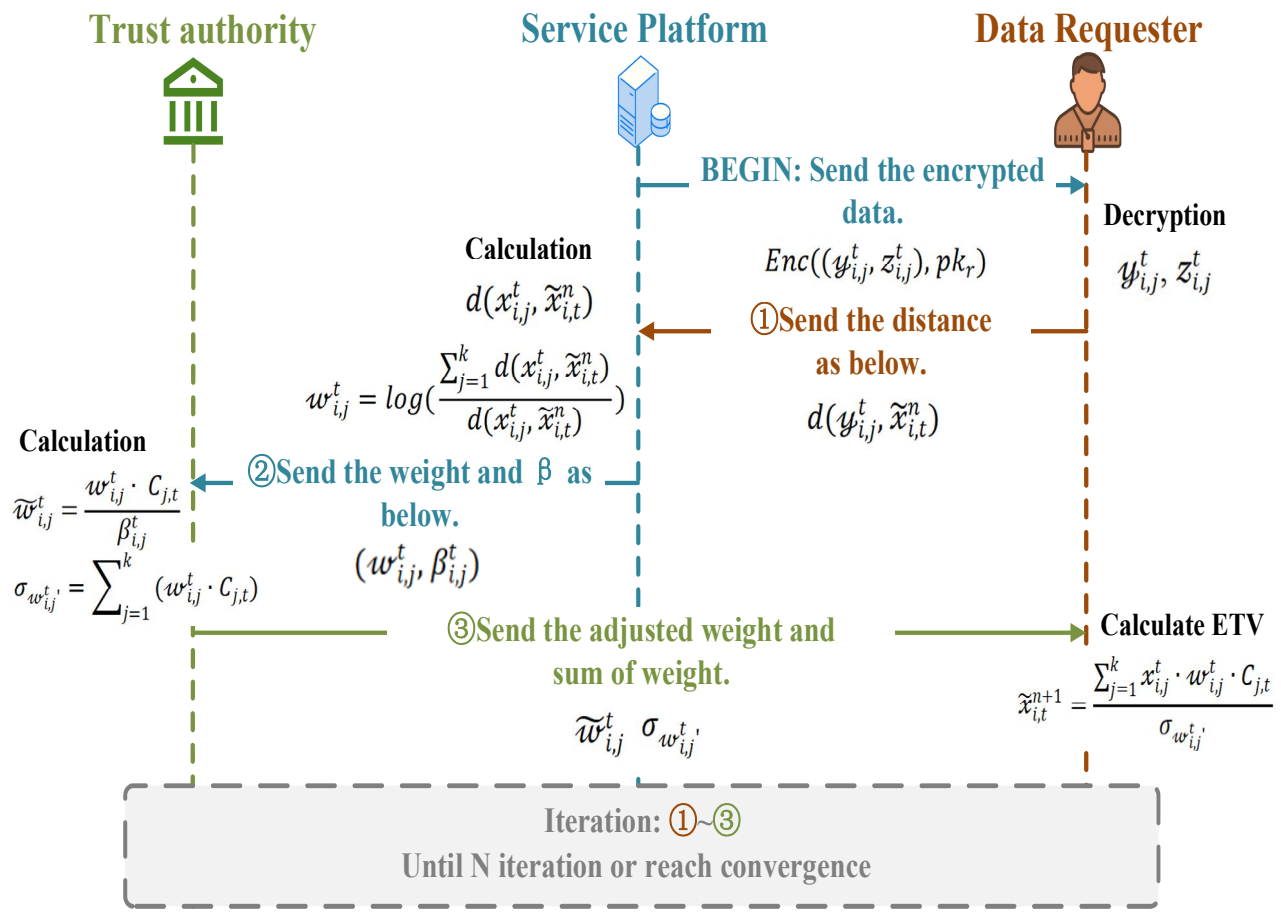

**Fig. 4.** The workflow of RTD-based data aggregation.

(1) Privacy of worker attributes in RDPP approach: RDPP ensures that worker attributes remain private while enabling reliable verification.

**Theorem 1**. Worker attributes $\mathcal{D}^t_{i,j}$ is protected.

*Proof.* The proof relies on the cryptographic properties of the encryption scheme used in RDPP approach. $C_{j,t}$ in worker attributes $\mathcal{D}^t_{i,j} = \{\mathcal{T}^t_{i,j}, \mathcal{L}^t_{i,j}, B^t_{i,j}, C_{j,t}\}$ is privacy-protected since $C_{j,t}$ is stored in the TA's database, and the TA is considered to be trustworthy. Each worker generates vector $\mathcal{v}_{\mathcal{D}^t_{i,j}} = \{\mathcal{v}_{\mathcal{T}^t_{i,j}}, \mathcal{v}_{\mathcal{L}^t_{i,j}}, \mathcal{v}_{B^t_{i,j}}\}$ as shown in Eq. (16) and encrypts $\mathcal{v}_{\mathcal{D}^t_{i,j}}$ according to Eq. (17). SP received $E_{\mathcal{K}}(\mathcal{v}_{\mathcal{D}^t_{i,j}})$ but is unable to derive the original attribute values because the encryption process $E_{\mathcal{K}}(\mathcal{v}_{\mathcal{D}^t_{i,j}})$ is a one-way function meaning it is computationally infeasible to reverse the process without knowing the encryption matrix $\mathcal{K}$ and the random value $\tau^t_{i,j}$. Even if an attacker attempts a brute-force attack, the security of the encryption is protected by using large-bit length $\tau^t_{i,j}$ and varying the encryption matrix $\mathcal{K}$, making it practically impossible to crack. ☐

(2) Effectiveness of worker recruitment in RDPP approach: RDPP ensures that only workers whose attributes satisfy the task requirements can be recruited.

**Theorem 2**. Recruitment results $\forall\psi^t_{i,j} = 1, \mathcal{D}^t_{i,j}$ satisfy $\mathcal{D}_{i,t}$.

*Proof.* For $C_{j,t}$ in worker attributes $\mathcal{D}^t_{i,j}$, $r^t_{i,j} = 1$ only if $C_{j,t} \geq C_{i,t}$. According to Eq. (30), Eq. (31), $\forall\psi^t_{i,j} = 1$, $C_{j,t} \geq C_{i,t}$.

$$(r^t_{i,j} = 1) \Rightarrow C_{j,t} \geq C_{i,t} \tag{30}$$

$$(\psi^t_{i,j} = 1) \Rightarrow (r^t_{i,j} = 1) \tag{31}$$

The verification process ensures that the selected worker's remain attributes fall within the specified intervals. Since the attributes $\mathcal{T}^t_{i,j}, \mathcal{L}^t_{i,j}, B^t_{i,j}$ are treated equivalently in the verification process, we focus on the time attribute $\mathcal{T}^t_{i,j}$ for simplicity. Therefore, the proof is simplified to $\forall r^t_{i,j} = 1, \mathcal{T}^t_{i,j}$ satisfy $\mathcal{T}_{i,t}$. The $[\mathrm{A} \cdot \mathrm{B}]$ operation is computed as the matrix multiplication of matrix A and the transpose of matrix B. As shown in Eq. (16) and Eq. (20), Eq. (21) can be derived as follows:

$$\begin{aligned}
&\left[E_{\mathcal{K}}(\mathcal{v}_{\mathcal{T}^t_{i,j}}) \cdot E_{\mathcal{K}}(d_{\mathcal{T}^t_{i,j}})\right] \\
&= \left[(\tau^t_{i,j} \cdot |det(\mathcal{K})| \cdot \mathcal{K}^{-1} \cdot \mathcal{v}_{\mathcal{T}^t_{i,j}}) \cdot (\tau^{t\,'}_{i,j} \cdot \mathcal{K}^T \cdot d_{\mathcal{T}^t_{i,j}})^T\right] \\
&= \tau^t_{i,j} \cdot \tau^{t\,'}_{i,j} \cdot |det(\mathcal{K})| \cdot \mathcal{K}^{-1} \cdot \mathcal{K} \cdot \mathcal{v}_{\mathcal{T}^t_{i,j}} \cdot d_{\mathcal{T}^t_{i,j}}{}^T \\
&= \tau^t_{i,j} \cdot \tau^{t\,'}_{i,j} \cdot |det(\mathcal{K})| \cdot \mathcal{v}_{\mathcal{T}^t_{i,j}} \cdot d_{\mathcal{T}^t_{i,j}}{}^T \\
&= \tau^t_{i,j} \cdot \tau^{t\,'}_{i,j} \cdot |det(\mathcal{K})| \cdot (\mathcal{T}^t_{i,j} - \overleftarrow{t_{i,t}}) \cdot (\overrightarrow{t_{i,t}} - \mathcal{T}^t_{i,j}) \cdot (\mathcal{T}^t_{i,j} + \varepsilon)
\end{aligned} \tag{32}$$

Where $\tau^t_{i,j}$ and $\tau^{t\,'}_{i,j}$ are randomly generated positive integers, $|det(\mathcal{K})|$ is the determinant of the encryption matrix $\mathcal{K}$, and $\varepsilon$ is a publicly known $\mu$-bit positive number, which ensures that $(\mathcal{T}^t_{i,j} + \varepsilon)$ is always positive. We consider all cases based on the position of $\mathcal{T}^t_{i,j}$ relative to the interval $[\overleftarrow{t_{i,t}}, \overrightarrow{t_{i,t}}]$.

Case 1: $\mathcal{T}^t_{i,j}$ falls to the left of the interval, i.e., $\mathcal{T}^t_{i,j} - \overleftarrow{t_{i,t}} < 0$. In this case, $(\mathcal{T}^t_{i,j} - \overleftarrow{t_{i,t}}) < 0$ and $(\overrightarrow{t_{i,t}} - \mathcal{T}^t_{i,j}) > 0$. Since $(\mathcal{T}^t_{i,j} - \overleftarrow{t_{i,t}})$ is negative while all other terms are positive, the outcome is negative, violating the verification condition.

Case 2: $\mathcal{T}^t_{i,j}$ falls to the right of the interval, i.e., $\mathcal{T}^t_{i,j} - \overrightarrow{t_{i,t}} > 0$. In this case, $(\mathcal{T}^t_{i,j} - \overleftarrow{t_{i,t}}) > 0$ and $(\overrightarrow{t_{i,t}} - \mathcal{T}^t_{i,j}) < 0$. Since $(\overrightarrow{t_{i,t}} - \mathcal{T}^t_{i,j})$ is negative while all other terms are positive, the outcome is negative, violating the verification condition.

Case 3: $\mathcal{T}^t_{i,j}$ falls within the interval, i.e., $(\mathcal{T}^t_{i,j} - \overleftarrow{t_{i,t}}) > 0$ and $(\overrightarrow{t_{i,t}} - \mathcal{T}^t_{i,j}) > 0$. Since all terms are positive, the outcome is positive, satisfying the verification condition.

As all cases above, RDPP ensures that only workers whose attributes satisfy the task requirements can be recruited. ☐

(3) Security of perturbed data: In RDPP approach, both SP and worker are aware of the perturbation factors $(\alpha^t_{i,j}, \beta^t_{i,j})$. SP could potentially derive the data collected by workers from the perturbed data $y^t_{i,j}$, $z^t_{i,j}$. To prevent this, the perturbed data is encrypted using the DR's public key $pk_r$, which ensures the integrity and confidentiality of data collected by workers.

**Theorem 3**. $Enc((y^t_{i,j}, z^t_{i,j}), pk_r)$ can be decrypted only by DR.

*Proof.* The encryption of $y^t_{i,j}, z^t_{i,j}$ employs the ElGamal encryption algorithm, which lies on the computational infeasibility of the discrete logarithm problem (DLP) in the multiplicative group $\mathbb{Z}^*_{p}$. Specifically, given a large prime $p$, a generator $c$ of $\mathbb{Z}^*_{p}$, and the public key $pk_r = c^{sk_r} \bmod p$, it is computationally infeasible to determine the private key $sk_r$ from $pk_r$ without solving the DLP.

Given a ciphertext $(\mathbb{c}_1, \mathbb{c}_2)$ where $\mathbb{c}_1 = c^k \bmod p$ and $\mathbb{c}_2 = m \cdot pk_r{}^k \bmod p$, an attacker would need to solve the DLP to determine the random integer $k$ or the private key $sk_r$. Since solving the DLP is considered computationally infeasible for sufficiently large $p$, the perturbed data remains secure against attacks. This ensures that only DR, who possesses the private key $sk_r$, can decrypt and access the data. □

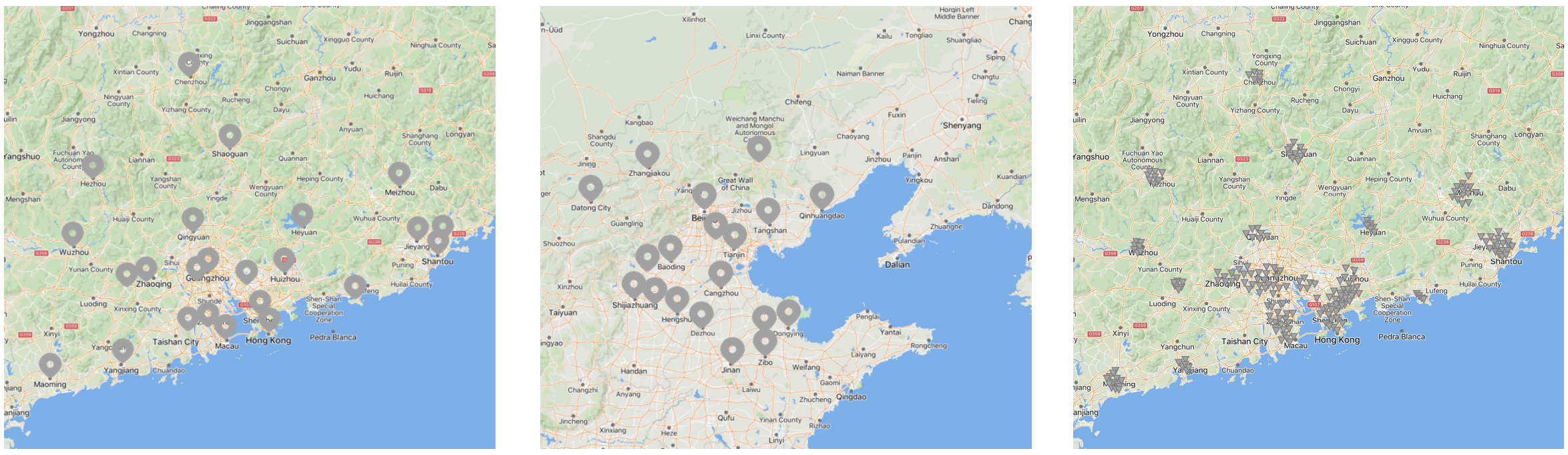

(a)24 cities around Guangzhou (b)19 cities around Beijing (c)Generated tasks

**Fig. 5.** City Distribution and Task Generation

(4) Confidentiality of data collected by workers: In RTD-based data aggregation, we thoughtfully design the data aggregation process to protect the confidentiality of the data, preventing data misuse.

**Theorem 4**. Data collected by worker remains confidential throughout the entire process.

*Proof.* The confidentiality of the worker's data $x_{i,j}^t$ is ensured through the perturbation mechanism. We proof the validity of **Theorem 4** from the viewpoints of SP and DR respectively.

DR's perspectives: after decrypting the received data using DR's private key $sk_r$, DR obtains perturbed data as Eq. (10). Since DR is unable to obtain perturbation factors $(\alpha_{i,j}^t, \beta_{i,j}^t)$, it fails to isolate the exact values of worker's data $x_{i,j}^t$.

SP's perspectives: SP, who shares perturbation factors $(\alpha_{i,j}^t, \beta_{i,j}^t)$ with the workers, receives $d(y_{i,j}^t, \tilde{x}_{i,t}^n)$ from DR. Then, SP computes as Eq. (23) and obtains:

$$x_{i,j}^t - \tilde{x}_{i,t}^n \tag{33}$$

Since SP is unable to know n$^{\text{th}}$ iteration outcome $\tilde{x}_{i,t}^n$, which is initialized to a non-zero value, SP fails to figure out $x_{i,j}^t$. □

(1) Effectiveness of Data Aggregation: In RTD-based data aggregation, DR can obtain aggregated outcome without knowing the exact values of the worker's data.

**Theorem 5**. DR can obtain the aggregated outcome.

*Proof.* Upon receiving $\tilde{w}_{i,j}^t$ and $\sigma_{w_{i,j}^t}{}'$ from TA, DR multiplies $z_{i,j}^t$ with the adjusted weight $\tilde{w}_{i,j}^t$ as follows:

$$z_{i,j}^t \cdot \tilde{w}_{i,j}^t = x_{i,j}^t \cdot \beta_{i,j}^t \cdot \frac{w_{i,j}^t \cdot C_{j,t}}{\beta_{i,j}^t} = x_{i,j}^t \cdot w_{i,j}^t \cdot C_{j,t} \tag{34}$$

Then, DR sum the weighted data and divide by the total weight $\sigma_{w_{i,j}^t}{}'$ to obtain the aggregated result for the current iteration as shown in Eq. (28). □

## V. Performance Analysis

### *A. Experiment Overview*

In this section, we will introduce the datasets selected for the experiment, the experimental setup, and the specific analysis of the experiment. We compare the advantages of the RDPP-TD scheme over existing solutions from multiple perspectives.

### *B. Dataset Introduction*

This experiment uses a dataset from Microsoft Research's "Urban Air" project, which includes air quality, meteorological, and weather forecast data. As shown in Fig. 5(a) and Fig. 5(b), the dataset covers four major cities in China—Beijing, Tianjin, Guangzhou, and Shenzhen—as well as their surrounding 39 cities, spanning the period from May 1, 2014, to April 30, 2015. This comprehensive dataset is particularly well-suited for research in MCS, especially for applications involving task allocation and data collection that rely on time and geographical location.

(1) Key features of the dataset: the dataset used in this experiment features high temporal and spatial resolution, with meteorological data (e.g., temperature) recorded hourly and accompanied by latitude and longitude information, enabling precise task allocation based on time and location. Additionally, the dataset is sourced from real-world monitoring stations, providing a reliable benchmark for evaluating MCS systems.

(2) Application of the dataset: as shown in Fig. 5(c), the dataset is used to enhance MCS systems by optimizing task allocation and simulating real-world data collection. For task allocation, the system screens workers based on their time and location to ensure efficient task distribution. For data collection, workers gather data at specified times and locations, replicating real-world processes.

### *C. Experimental Setup*

The simulation experiments are conducted on a system equipped with an AMD Ryzen 7 6800H processor with Radeon Graphics, running at 3.20 GHz and paired with 16 GB of RAM.

Table III. the generation of the workers.

| Worker category | Proportion | TRV |
|---|---|---|
| MWs | 30% | (0, 0.2) |
| AWs | 50% | [0.2, 0.7) |
| TWs | 20% | [0.7, 1) |

(1) Generation of workers: the true reputation value (TRV) of workers is generated based on three categories: malicious workers (MWs), average workers (AWs), and trust workers (TWs). The worker pool is divided into these three categories according to their respective proportions within the total number of workers, which is denoted as $N$. As shown in Table III, MWs, which account for 30% of the worker pool, are assigned TRV randomly within the range $(0, 0.2)$. AWs, making up 50% of the pool, have their TRV randomly assigned between $[0.2, 0.7)$. The remaining 20% are TWs, who are given TRV in the range [0.7, 1). This distribution ensures a diverse mix of worker reliability levels, reflecting a realistic scenario.

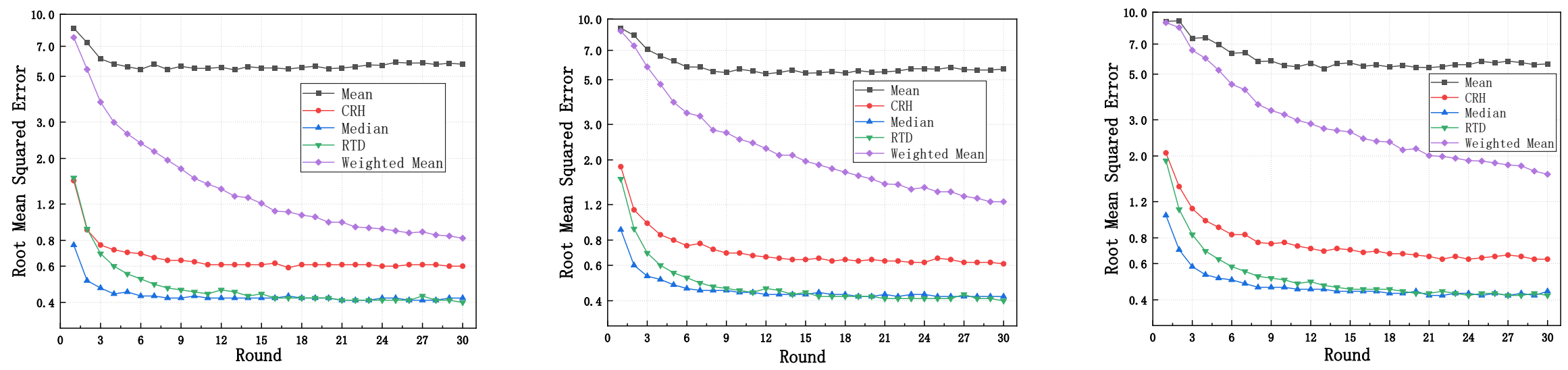

(a)$N$ = 200 (b)$N$ = 400 (c)$N$ = 600

**Fig. 6.** RMSE analysis with varied worker pool sizes.

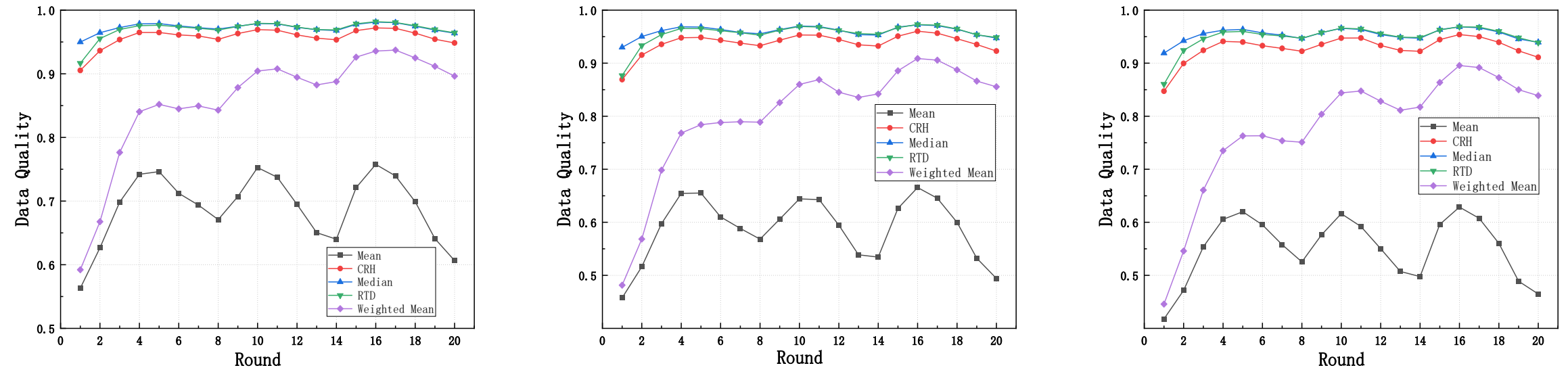

(a)$\lambda$ = 10 (b)$\lambda$ = 30 (c)$\lambda$ = 50

**Fig. 7.** Data quality analysis with varied λ.

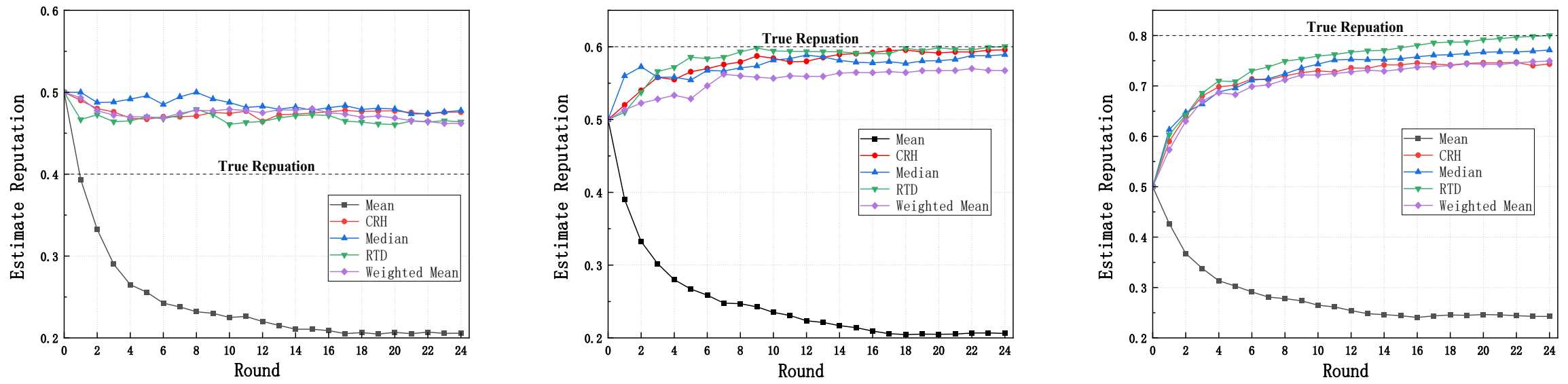

(a)$C_{j,t}$ = 0.4 (b)$C_{j,t}$ = 0.6 (c)$C_{j,t}$ = 0.8

**Fig. 8.** Reputation recognition analysis.

(2) Generation of data by workers: the data collected by workers are simulated by adding noise to the true values of weather indicators (e.g., temperature, air quality) from the dataset. Data quality depends on the worker's TRV $C_{j,t}$. TWs generate high-quality data with probability equals to $C_{j,t}$ and introduce larger noise with probability equals to $(1 - C_{j,t})$, while AWs behave similarly, producing better-quality data with probability equals to $C_{j,t}$ and larger noise with probability equals to $(1 - C_{j,t})$. In contrast, MWs generate data randomly within a wide range, independent of the true value, simulating unreliable data collection. This design reflects real-world scenarios where the reputation of worker impacts data quality, with higher TRV corresponding to more accurate data and lower TRV leading to significant errors.

(3) Performance evaluation metrics: this paper uses RMSE with GTD and Data Quality to evaluate the performance of data aggregation methods.

- RMSE with GTD: as mentioned in Eq. (1), $\Re$ measures the square root of the average of squared differences between the ETV and the GTD across all tasks. A lower $\Re$ indicates higher precision and better performance.
- Data Quality: as defined in Eq. (5), $Q$ evaluates the overall quality of ETV. Higher $Q$ values indicate better data quality and more reliable data aggregation.

### *D. Experimental Analysis*

In this subsection, we conduct a detailed experimental analysis, focusing on four key aspects: the accuracy of ETV, the performance of reputation estimation, the robustness under collusion attacks, and the ablation analysis. These analyses comprehensively demonstrate the effectiveness and robustness of the RDPP-TD scheme.

#### *1). Accuracy of ETV*

We compared RTD with four other data aggregation methods (mean, median, weighted mean, and CRH) using RMSE and Data Quality as evaluation metrics.

RMSE evaluation is designed with a fixed task count of $1500$, divided into rounds of $50$ tasks each under varying initial worker pool sizes $N$. As shown in Fig. 6, RTD and the median method exhibited similar RMSE performance, both achieving a 33.3% improvement over CRH. Notably, the performance of weighted mean and mean is significantly inferior. As the initial worker pool size $N$ increased, all data aggregation methods require more rounds to stabilize their RMSE, indicating that larger worker pools introduce greater complexity in achieving stable estimates.

Data Quality is performed with a fixed task count of $1000$, divided into rounds of $50$ tasks each. We investigated the impact of varying the parameter $\lambda\,(10, 20, 30)$ on Data Quality. Similar to the RMSE results, RTD and the median method demonstrate superior data quality performance, followed by CRH, weighted mean, and mean according to Fig. 7. As $\lambda$ increased, Data Quality became more sensitive to $q_{i,t}$ in Eq. (4), leading to a faster decline in performance for less accurate method.

*2). Performance of reputation estimation*

We compared the performance of various data aggregation methods in estimating workers' TRV, especially their ability to identify workers with unknown reputation. We created workers with TRV of 0.4, 0.6, and 0.8 and had them participate in numerous tasks. As Fig. 8 shows, by comparing the differences between the estimated reputation values and TRV under different methods after certain rounds, we found that the data aggregation method with RTD performed best in evaluating the reputation of these three types of workers, surpassing the other four methods. When TRV was 0.6 and 0.8, the performance of CRH and median methods varied, with the weighted mean method being the next best. In contrast, the estimated reputation values with mean significantly deviated from the TRV. In MCS with a reputation mechanism, the superior accuracy of RTD in estimating the reputation of workers contributes to enhancing the system's overall efficiency and performance.

*3). Robustness under collusion attacks*

In practical MCS systems, various complex scenarios may be encountered, and being subjected to collusive attacks by MWs is an important issue. Traditional data aggregation methods (e.g., CRH) show good robustness against data poisoning by non-colluding MWs, even when the proportion of MWs is high. However, when MWs collude to launch collusive attacks, they can succeed with at most just over half the workers in a task to fraudulently obtain task rewards. Therefore, how to effectively defend against collusive attacks has become a key focus for enhancing the security and robustness of MCS systems.

We firstly analyzed the RMSE and Data Quality of five data aggregation methods under different task worker counts $k \in \{10, 20, 30\}$ and MWs proportions $\theta \in \{0.2, 0.4, 0.6, 0.8\}$. Results from Fig. 9 show that at MWs proportions of 0.2 and 0.4, RTD and the mean methods outperformed others in RMSE and Data Quality, both achieves a 62.5% improvement in data accuracy compared to CRH. CRH performed well at 0.2 but dropped significantly at 0.4. The mean method showed the poorest performance. As the MWs proportion rose to 0.6, RMSE increased and Data Quality decreased for all methods.

Table VI. the value of the parameter.

| Parameter name | Values |
|---|---|
| $N$ | 200, 400, 600 |
| $\lambda$ | 10, 20, 30 |
| $C_{j,t}$ | 0.4, 0.6, 0.8 |
| $k$ | 10, 20, 30 |
| $\xi$ | 0.3, 0.4, 0.5 |
| $C_{i,t}$ | 0.3, 0.4, 0.5 |
| $\theta$ | 0.2 0.4 0.6 0.8 |

However, RTD demonstrated exceptional stability and superiority, with the smallest performance change. At an MWs proportion of 0.8, the performance of RTD significantly outpaced the other four methods. The improvement in data accuracy reaches 90.5% compared to CRH, maintaining robustness and accuracy. The weighted mean method, which calculates weights based on estimated worker reputation, provided some defense against high MWs proportions and ranked second. In contrast, CRH, median, and mean had RMSE over 15 and Data Quality below 0.4 at higher MWs proportions, indicating failure to resist collusion attacks. Increasing $k$ also led to performance drops for all methods at MWs proportions of 0.6 and 0.8.

We further explored how changes in the parameter $\lambda$ and the reputation values of MWs $\xi$ affect data quality of RTD. We tested $\lambda$ values of 10, 20, 30, and 40, and set the $\xi$ to 0.2, 0.3, 0.4, and 0.5. The analysis covered scenarios where MWs made up 0.4, 0.6, and 0.8 of the task workforce. From Fig. 10, it's clear that Data Quality declines as $\lambda$ and $\xi$ increase. A higher $\lambda$ makes the system more sensitive to errors, while a larger reputation range for MWs higher their weights in the aggregation process. When MWs account for 0.4 of the task workforce, the impact on data quality is minimal, as the MWs' reputation increases from 0.2 to 0.5, the data quality decreases from 0.9756 to 0.9126, a decrease of approximately 6.46%. However, as $\xi$ increases, the impact becomes more significant, and Data Quality drops noticeably. When MWs account for 0.8 of the task workforce, as the reputation of MWs increases from 0.2 to 0.5, the Data Quality decreases from 0.820 to 0.4128, a decrease of approximately 49.67%. This is because the larger reputation range grants MWs more influence in the RTD aggregation process.

In summary, RTD outperforms four other methods in resisting collusion attacks and significantly boosts the robustness of MCS systems. Compared to the CRH method, RTD improves Data Quality from 62.5% to 90.5% as the proportion of MWs in the worker set increases from 0.2 to 0.8. Nevertheless, its performance declines when facing collusion attacks from MWs with higher reputation values. To address this, the RDPP-TD scheme integrates the advantages of RTD in data aggregation and worker reputation estimation, and combines it with the RDPP approach, which has a worker screening function.

*4). Ablation analysis*

As previously mentioned, the RDPP-TD scheme integrates the RDPP approach for secure reliability verification and the RTD

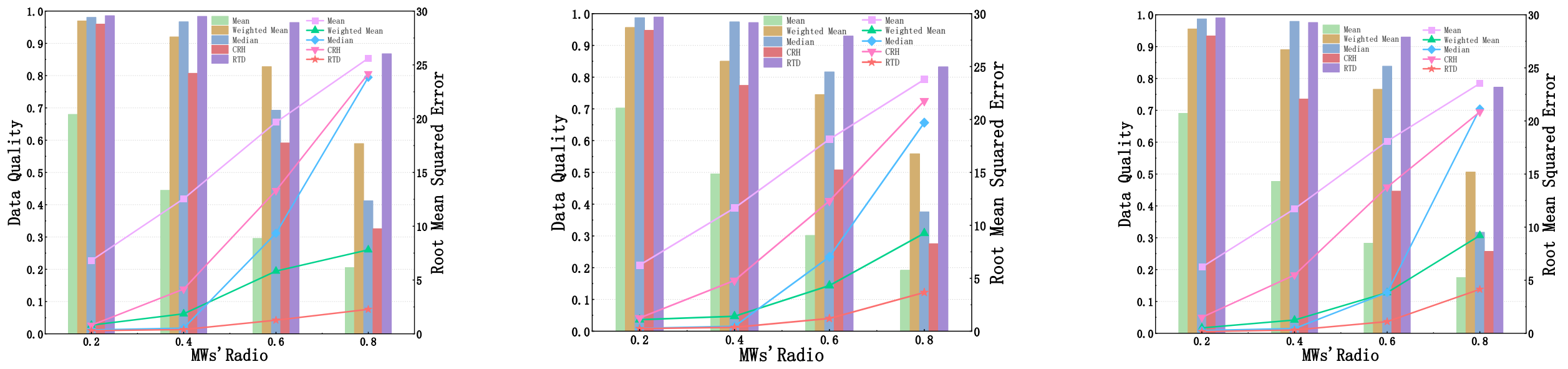

(a)$k$ = 10 (b)$k$ = 20 (c)$k$ = 30

**Fig. 9.** Robustness against collusion attacks.

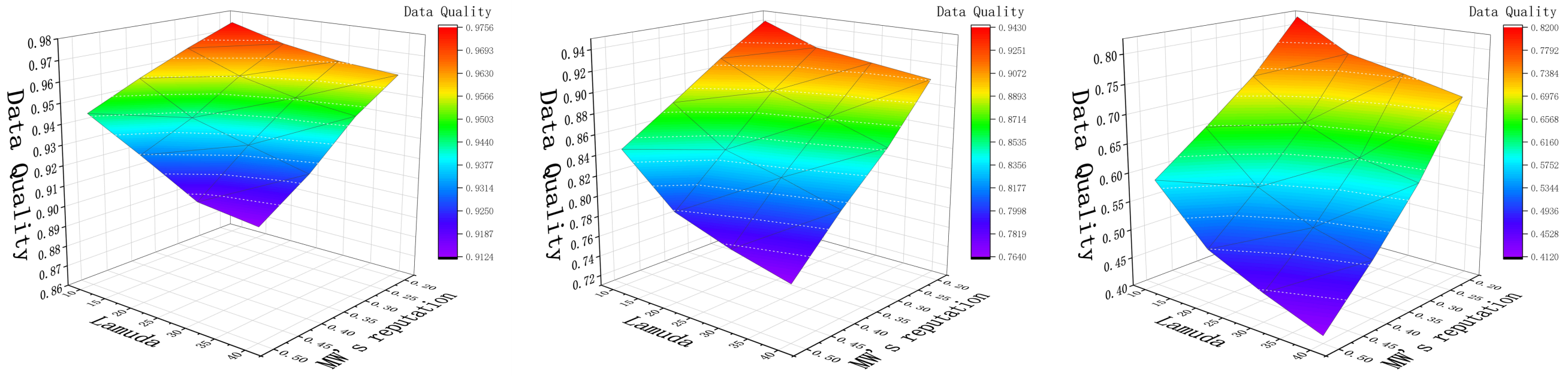

(a)$\theta$ = 0.4 (b)$\theta$ = 0.6 (c)$\theta$ = 0.8

**Fig. 10.** Data quality analysis of RTD.

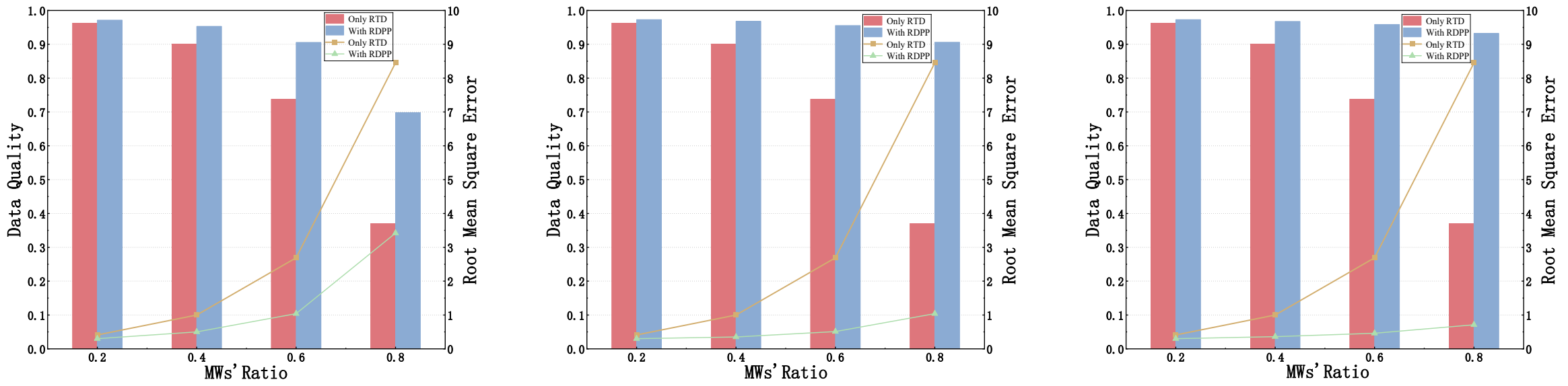

(a)$C_{i,t}$ = 0.3 (b)$C_{i,t}$ = 0.4 (c)$C_{i,t}$ = 0.5

**Fig. 11.** Ablation analysis of RDPP approach.

method for accurate and robust data aggregation. The RDPP approach's ability to screen worker reliability is crucial for enhancing subsequent data aggregation performance and the system's robustness against collusion attacks. We conducted an ablation analysis on the RDPP-TD scheme to compare the performance differences between standalone RTD and RTD with RDPP under varying conditions. Specifically, we examined scenarios where the reputation requirement $C_{i,t}$ in the RDPP approach was set to 0.3, 0.4, and 0.5, with MWs having a reputation range of 0.5 and comprising 0.2, 0.4, 0.6, and 0.8 of the workforce.

From Fig. 11, it is evident that at lower MWs proportions (0.2 and 0.4), RTD with RDPP shows modest improvements of less than 5.8% over RTD alone. However, as the MWs proportion increases to 0.6, the performance gap widens significantly. RTD with RDPP maintains a lower RMSE and a Data Quality above 0.9, whereas RTD's RMSE increases markedly, and its Data Quality drops below 0.8, resulting in a 22.7% improvement for RTD with RDPP. When the MWs proportion reaches 0.8, RTD's performance degrades further. Even though RTD outperforms four other methods in this scenario, RTD with RDPP demonstrates markedly superior robustness, achieving an 88.6% improvement in Data Quality. RTD with RDPP significantly outperforms standalone RTD in both RMSE and Data Quality, highlighting the advantages of combining RDPP and RTD in defending against collusion attacks. Furthermore, as the reputation threshold in RDPP increases from 0.3 to 0.4 and 0.5, the RMSE of RTD with RDPP stabilizes around 1, and the Data Quality remains above 0.9, indicating enhanced performance. Specifically, when the reputation threshold reaches 0.5, the Data Quality improves by 152%. This means that MCS systems can dynamically adjust the RDPP reputation threshold to select higher-quality workers. In the face of sustained collusion attacks, increasing the RDPP screening criteria can further strengthen the system's robustness and improve the accuracy of ETV. Overall, the RDPP-TD scheme not only leverages the strengths of RTD but also significantly enhances the system's resilience against sophisticated attacks through the integration of the RDPP approach.

*5). Experimental results*

Through comprehensive and diverse performance evaluation, the proposed RTD method achieves a 33.3% improvement in aggregation accuracy compared to the conventional CRH approach. Specifically, RTD demonstrates superior performance in worker reputation identification, with statistically significant enhancements in precision. Under adversarial scenarios involving high-proportion MWs collusion attacks, RTD exhibits significantly enhanced robustness, maintaining system efficiency with a lower error rate than CRH under equivalent attack intensities. Specifically, as the proportion of MWs increases to 0.8, RTD achieves a 90.5% improvement in data accuracy, demonstrating its robustness under high-intensity collusion attacks.

The RDPP-TD scheme integrates the efficiency-optimized RTD method with the privacy-preserving RDPP worker screening approach. Experimental results demonstrate that RDPP-TD achieves over 62.5% improvement in computational efficiency over CRH while maintaining over 0.9 Data Quality under collusion attacks. This approach establishes a novel paradigm for secure and efficient MCS infrastructure, addressing both performance and privacy challenges in adversarial environments.

## VII. Conclusion and Further Works

In this paper, we propose the RDPP-TD scheme based on a reputation mechanism to address the challenges of worker screening, reputation identification, and ETV computation under privacy protection. Within the RDPP-TD scheme, we design the RTD data aggregation method, which integrates a reputation mechanism into the data aggregation process, adjusting worker weights based on their historical task performance. The RTD data aggregation method enhances the accuracy of ETV and the robustness of MCS under collusion attacks. Furthermore, we introduce the RDPP approach, which leverages ElGamal cryptographic algorithms, matrix encryption, and perturbation factors to achieve comprehensive privacy protection for workers. This paper provides a detailed introduction to the processes and key algorithms of the RDPP-TD scheme. Our theoretical analysis demonstrates that the proposed scheme ensures rationality, security, and privacy. In the performance analysis, we assessed the effectiveness and superiority of the proposed method by comparing the CRH, median, weighted mean, and mean methods. The experimental results demonstrated that our method exhibited superior performance.

In future work, we will prioritize enhancing the efficiency of data quality and worker screening approaches. Sustained efforts will be directed toward improving the system's capability to recognize workers. Additionally, we will further investigate the responsibility allocation of the SP, TA and refine the workflow integration between the SP, TA, DR, and workers, with the aim of designing a more efficient and collaborative system architecture.

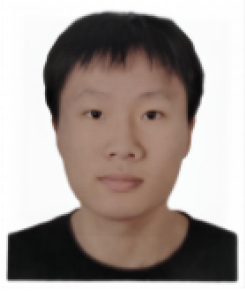

**Lijian Wu** is currently a student at the School of Computer Science and Engineering, Central South University, China. His research interests include mobile crowdsourcing and artificial intelligence. Email: 8208221106@csu.edu.cn.

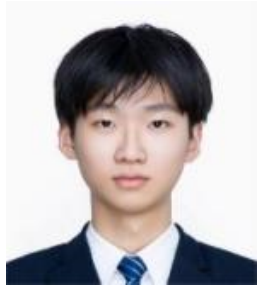

**Weikun Xie** is currently a student at the School of Information Security, Central South University, China. His research interests include Mobile Crowdsensing, System Security and Cyber Security. E-mail: xieweikun@csu.edu.cn.

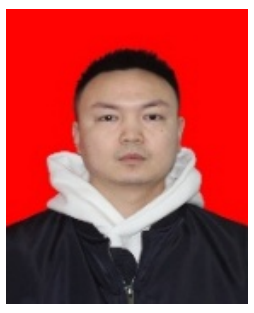

**Wei Tan** is currently working toward the PhD degree with the School of Computer Science and Engineering, Central South University, China. His research interests include mobile crowd sensing, edge computing and incentive mechanism. Email: wei.tan@csu.edu.cn.

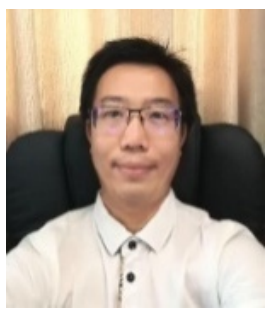

**Tian Wang** received the BSc and MSc degrees in computer science from Central South University, in 2004 and 2007, respectively, and the PhD degree in computer science from the City University of Hong Kong, in 2011. Currently, he is a professor with the Institute of Artificial Intelligence and Future Networks, Beijing Normal University. His research interests include Internet of Things, edge computing, and mobile computing. E-mail: tianwang@bnu.edu.cn.

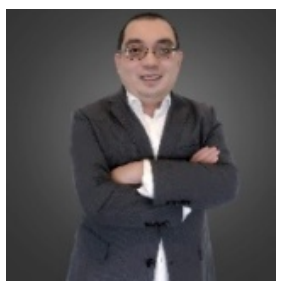

**Houbing Herbert Song** (M'12–SM'14-F'23) received the Ph.D. degree in electrical engineering from the University of Virginia, Charlottesville, VA, in August 2012. He is currently a Tenured Associate Professor at University of Maryland, Baltimore County (UMBC), Baltimore, MD. Prior to joining UMBC. His research interests include cyber-physical systems/internet of things. E-mail: h.song@ieee.org.

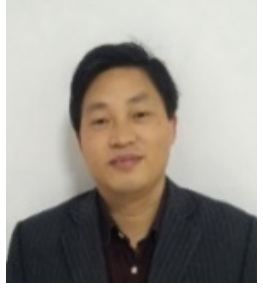

**Anfeng Liu** received the M.Sc. and Ph.D. degrees from Central South University, China, in 2002 and 2005, respectively, both in computer science. He is a Professor with School of Information Science and Engineering of Central South University, China. His major research interests are Internet of Things and crowdsourcing. e-mail. afengliu@mail.csu.edu.cn.